\documentclass[aps,pra,twocolumn,groupedaddress, showpacs]{revtex4}
\usepackage{graphicx}
\usepackage{graphics}
\usepackage{color}      
\usepackage{subfigure}

\begin{document}


\definecolor{AsGreen}{rgb}{0.3,0.8,0.3} 
\definecolor{AsRed}{rgb}{0.6,0.0,0.0}
\definecolor{AsLightBlue}{rgb}{0,0.6,0.6}
\definecolor{AsPurple}{rgb}{0.6,0,0.6}

\hyphenpenalty=5000
\tolerance=1000

\title{Long-range sound-mediated dark soliton interactions in trapped atomic condensates}
\author{A.~J. Allen}
\email{a.j.allen1@ncl.ac.uk}
\author{D.~P. Jackson}
\author{C.~F. Barenghi}
\author{N.~P. Proukakis}
\affiliation{School of Mathematics and Statistics, 
Newcastle University, Newcastle upon Tyne, NE1 7RU, United Kingdom}
\date{\today}
\begin{abstract}
A long-range soliton interaction is discussed whereby two or more dark
solitons interact in an inhomogeneous atomic condensate, modifying their respective
dynamics via the exchange of sound waves without ever coming into direct
contact.  An idealized double well geometry is shown to yield perfect energy
transfer and complete periodic identity reversal of the two solitons.  Two
experimentally relevant geometries are analyzed which should enable the
observation of this long-range interaction.
\end{abstract}
\pacs{
67.85.-d,  
03.75.Lm, 
05.45.Yv.  
}
\maketitle
\section{Introduction}
Solitons are an important phenomenon present in many areas of nonlinear
physics \cite{solitons} arising from the balance between dispersion and nonlinearity. 
Dark solitons are localized waves propagating within a background medium, 
and are characterized by a dip of the density 
and a phase slip \cite{kivshar_review}.
They have recently become a topic of intense research in weakly-interacting atomic
Bose-Einstein condensates \cite{frantzeskakis_book}, being routinely generated in a number of experiments, both in a controlled manner \cite{burger,phillips,Becker,Weller,Stellmer,huletsols} and as a result of
dynamical processes \cite{Tik,dutton,engelsdyna,ketterlemerging,KZ}.

Recent experiments have been able to generate appropriate low-temperature conditions in quasi-1D geometries enabling the
observation of one or more (undamped) dark soliton oscillations in a 
harmonically-confined condensate \cite{Becker,Weller,Stellmer}
 as predicted at the mean field 
level \cite{busch_anglin,makarov,frantzeskakis2002}. 
In general, the presence of the axial confinement in these 
experiments breaks the integrability of the system, 
rendering the soliton unstable to sound emission along its axis of propagation, inducing it to decay \cite{busch_anglin,Leadbeater,pelinovsky}.
Earlier work by us has revealed the crucial role of the harmonicity 
of the underlying trap in
stabilizing the propagating soliton against decay at low temperatures 
via continuous cycles of soliton-sound interactions 
 \cite{Leadbeater,Parker2010}, when all other decay mechanisms
are suppressed for the timescales of interest
\cite{muryshev,fedichev,Carr,Feder,jbrand,mury,sacha,proukakis_analogies,jackson,cockburn,ruostekoski,quantumdecay}.

Here we show that soliton-sound interactions 
also play a key role in the dynamics of two or more solitons in an inhomogeneous atomic condensate, 
and we identify optimal realistic experimental conditions 
for observing such an effect.  
In particular, we analyse the motion of solitons oscillating 
within either a single harmonically-confined condensate or different
spatially-separated sub-regions of a condensate and conclude that,
within mean field theory, 
dark solitons can interact via the emission/absorption 
of sound waves over a long range, 
without ever approaching very close to each other.  We refer to this interaction mechanism 
as sound-mediated dark soliton 
interactions, and suggest an experiment that could probe such an effect.

Although this effect bears close analogies to earlier studies in various physical systems and
configurations (see below) \cite{foursa,makarov,frantzeskakis_condmat,quantumdecay,segev}, 
this study is distinct in that 
(i) the integrability of the system is lifted by the longitudinal inhomogeneous confinement,
as in current cold atoms experiments, and that
(ii) at sufficiently low temperatures such exchange need not necessarily lead to the decay of the
solitons involved, but it can instead lead to a periodic exchange of energy between them.
We wish to note here that {\it direct} collisions of two dark solitons were
previously studied both in a homogeneous \cite{foursa,makarov} 
and a harmonically confined medium \cite{frantzeskakis_condmat}, 
and where found to result at most in a phase shift of the emerging solitons.
Long-range dark soliton interactions have also been discussed in  \cite{quantumdecay},
in the context of an axially {\it homogeneous} condensate, in which solitons interact
via superradiance due to the quantised nature of the background fluctuations.
Finally, the proposed effect is reminiscent of long-range {\it bright} 
soliton interactions observed  in non-local nonlinear optical media 
over distances much larger that the soliton size
\cite{segev}.

\begin{figure}[b]
\vspace{0.25cm}
\includegraphics[scale=0.32]{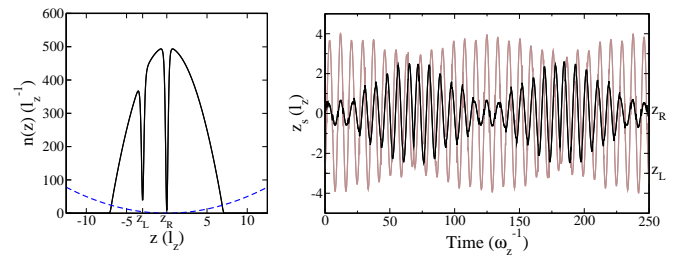}
\caption{(color online)  Motion of two non-identical dark solitons in a harmonically trapped condensate.  
Left: Density (black) and potential (dashed blue).  Right:  Position of left (solid, grey) and right soliton (solid, black) vs. time. 
(Initial parameters: $z_L=-3.0l_z$, $ v_s^L=0.3c$ and $z_R=0.0l_z$, $v_s^R=0.1c$, $\mu =
25\hbar\omega_z$).  }
\label{harm}
\end{figure}
    
\section{Demonstration of Effect}


\subsection{Two solitons within a single harmonic trap}

Previous work \cite{frantzeskakis_condmat,Weller} found that two identical, grey solitons displaced equally from the bottom of the harmonic trap and
travelling towards each other,  are reflected or pass through each other depending on their kinetic energies, 
indicating that
soliton-sound interactions have, at most, a minor effect on their mutual dynamics.
Although of very small magnitude, such modulations are actually present even in that context; however, this effect does become significant in the case of {\it non-identical}
solitons initially located at asymmetric locations within a single harmonic trap, as shown in Fig. 1.

As well-known, if there is just one soliton propagating in a harmonic trap, it 
perturbs the inhomogeneous background motion, setting up a dipole mode, but it also
periodically reabsorbs the sound emitted by its own motion \cite{Leadbeater}; 
the resulting oscillation exhibits a beating effect due to the different oscillation
frequencies of the soliton (=
$\omega_z/\sqrt{2}$)~\cite{proukakis_analogies,pelinovsky,makarov,Parker2010,busch_anglin,muryshev,fedichev,Leadbeater,frantzeskakis2002,schmelcher} and the induced background dipolar motion of the condensate ($=\omega_z$).  In the case of two non-identical solitons with different initial speeds,
the solitons are found to absorb sound and dissipate at different rates,
as each soliton re-interacts both with the sound it has emitted and with sound emitted by the other soliton.  This process, combined with periodic soliton collisions, leads to a much more pronounced modulation of the soliton oscillations than encountered in the case of a single soliton in a harmonic trap (whose corresponding amplitude
modulation would typically be a few percent of its initial amplitude \cite{parkerjphys}).

This effect is shown clearly 
in Fig.~\ref{harm} (Right):  
in particular, the fast soliton, initially imprinted away from the trap centre, gains energy, thereby 
exhibiting oscillations of decreasing amplitude during the initial time period, $t < 70 \omega_z^{-1}$.
This energy becomes available from the sound emission originating from the motion of the slower soliton, which in turn exhibits a small increase in its oscillation amplitude; at later times, the change in the relative phase of the two soliton oscillations leads to a reversal of the direction of the energy flow between them.
Clearly, however, their mutual energy exchange does not lead to a net decay within a single harmonic trap.



Our analysis is based on the one-dimensional (1D) Gross-Pitaevskii Equation
\begin{eqnarray}
i\hbar \frac{\partial \psi}{\partial t} = \left(- \frac {\hbar^2}{2m} \frac{\partial^2}{\partial z^2}+ V(z) + g |\psi|^2 -
\mu \right) \psi \;,
\label{eqn:gpe}
\end{eqnarray}
where $g$ parametrizes the effective 1D interatomic interaction via $ g =  ( 4\pi\hbar^2 a/m ) / 2 \pi
l_{\perp}^2$, where $a$
is the 3D s-wave scattering length;
here $l_{\perp} = \sqrt{\hbar/(m\omega_{\perp})}$ is the harmonic oscillator length in the tightly-confining transverse direction, of effective harmonic
frequency $\omega_{\perp}$, 
$m$ is the atomic mass and $\mu$ is the chemical potential.
For the trap parameters of a recent experiment used here \cite{Weller}, this correspond to an atom number of the order of $1000$ atoms.

\subsection{Two solitons in a spatially separated idealized double-harmonic trap}

To isolate the two effects, direct soliton collisions and soliton-sound interaction, in a manner that could also be achieved experimentally, we now move to a modified geometry which enables us to restrict the motion of the two solitons in two
spatially separated traps, while still allowing sound exchange between them.   Due to the known sensitivity of soliton-sound interaction on any trap anharmonicity \cite{Parker2010}, we initially choose an
idealized geometry $V(z) = m \omega_{z}^2 (z \pm z_0)^2 / 2$, 
with two minima at $\pm z_0$, such that the region
$z${\tiny ${ _{\geq}^{<}}$}$ 0$ provides a `harmonic environment' for each soliton individually; this
physical separation enables us to probe soliton-sound interactions over a long spatial range, and for long times.


The amount of
sound transferred between the two sides of the trap is set by the ratio of $V_0/\mu$,
where $V_0 = m \omega_z^2 z_0^2 /2$ is the height of the barrier between the neighbouring harmonic traps. 
Throughout this work, we fix this ratio to $V_0/\mu\approx 0.9$, 
which allows sound transfer between the two wells while keeping 
the solitons individually confined in different wells.  

\begin{figure}
\includegraphics[scale=0.3,angle=270]{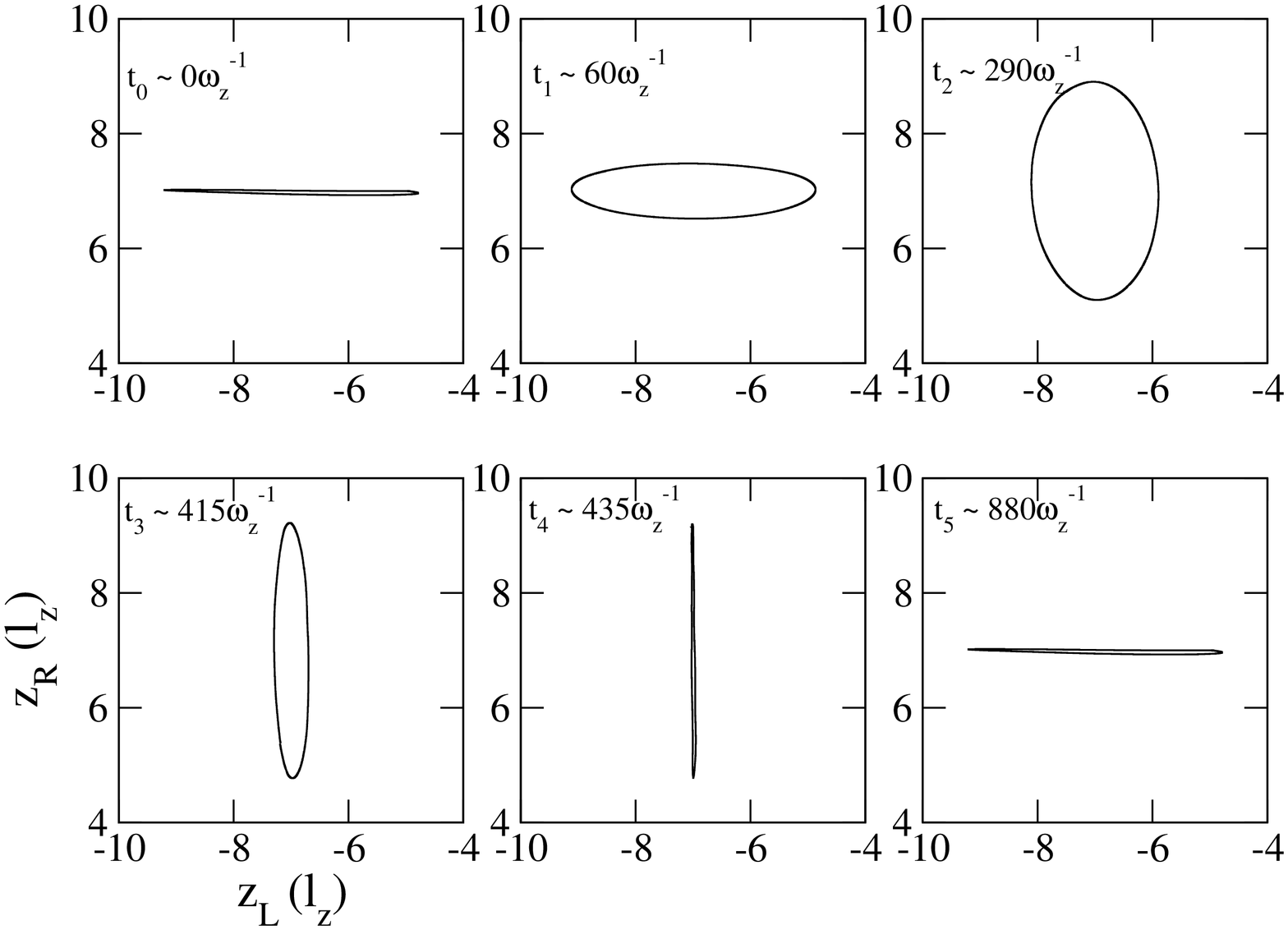}
\vspace{0.25cm}
\includegraphics[scale=0.25,angle=270]{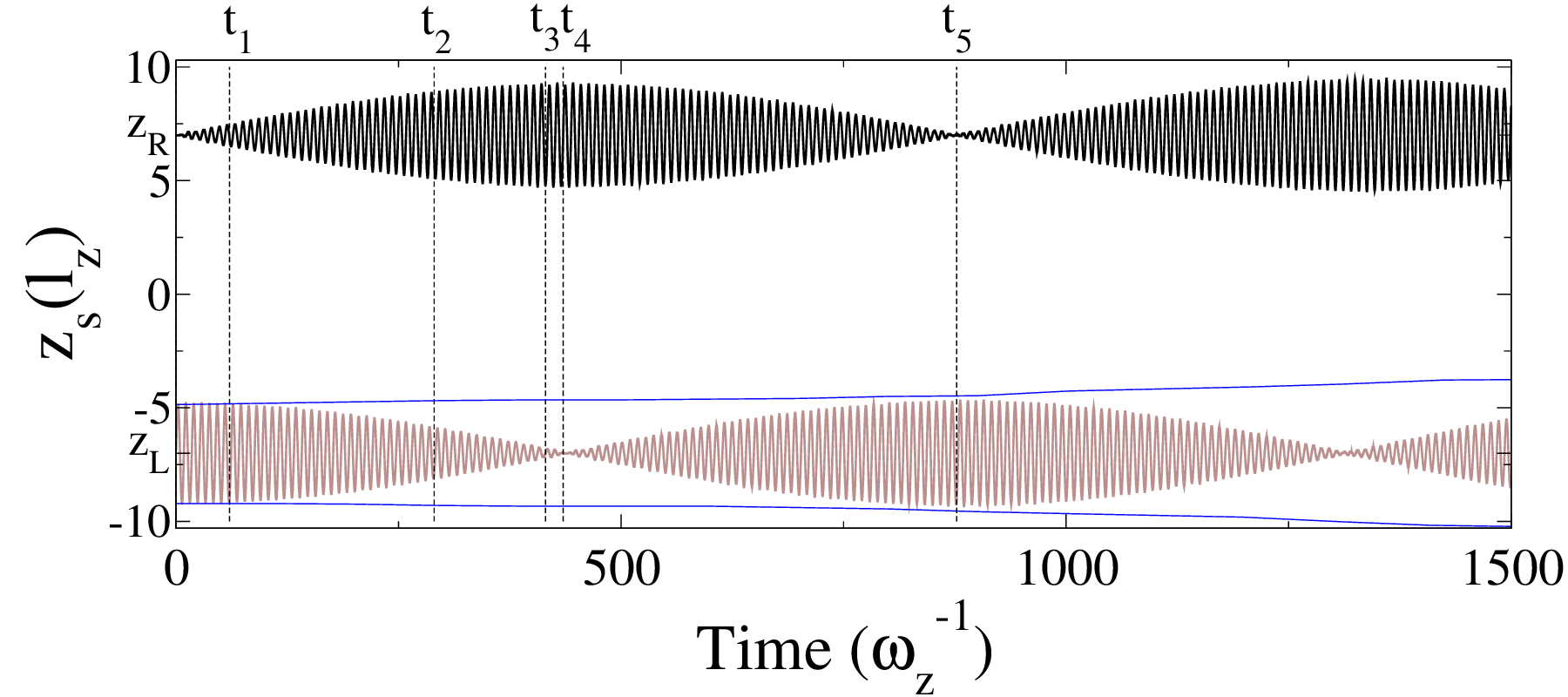}
\caption{(color online) 
Top images: Snapshots of soliton position in the left trap, $z_L$, against that in the right trap, $z_R$, at various times. 
Bottom: Oscillations of two solitons 
($v_s^L=0.3c$, $v_s^R=0$)
in an idealized double harmonic trap
( with $V_0/\mu =0.98$, $z_L = -7l_z$, $z_R=7l_z$),
showing the complete
interchange of the identities of the two solitons.  
Blue lines indicate the growing oscillation envelope of the initially
 $0.3c$ speed soliton placed in the left well when
there is no soliton in the other well.  
}
\label{doub}
\end{figure}

To monitor this idealized energy exchange mechanism, we consider for simplicity
a stationary black soliton in the centre of the right well ($z_R = 7l_z$), and a
moving, $0.3c$ (grey) soliton in the centre of the left well ($z_L=
-7l_z$), and track their subsequent coupled evolution. 
As the left soliton oscillates, it emits sound; while part of it remains
confined to the left well, thus re-interacting with the same soliton, the remaining part moves across the barrier into the other well (since $V_0 < \mu$).  The sound entering the right well
disturbs the initially stationary soliton present there, and causes it to move, itself radiating sound, part of which also becomes transferred into the left well.  
This process of emission, reabsorption and mutual sound exchange between the two solitons continues and results in a periodic reversal of their characteristics. 
This effect is clearly shown in Fig.~\ref{doub} (bottom).
One can think of one soliton acting as a parametric external driver for the other soliton~\cite{parametricdriving}, with the frequency of the driver set dynamically by soliton-sound interaction. 

Note that when the soliton in one well is moving the fastest, its amplitude of oscillation is at a maximum, and the soliton in the other well is stationary (its depth is maximum), and vice versa.  This transfer of energy is more apparent when plotting the position
of the soliton in the left well against the position of the soliton in the right well at different times of their coupled evolution, as shown in Fig.~\ref{doub} (top).  This figure corresponds to a temporal duration of approximately one soliton oscillation in the trap, i.e. $\tau = 2 \sqrt2\pi\omega_z^{-1}$; the bottom figure shows the individual soliton oscillations,
highlighting the times around which the top images were taken.  
Note that after a few more cycles the maximum amplitude of oscillation will gradually start increasing, resulting in an imperfect energy transfer;
eventually
both solitons will decay (even within pure mean field theory), due to the anharmonicity in the region of the barrier.  Increasing $V_0/\mu$ will result in observation of the identity
reversal over a much longer period of time, conversely, reducing the ratio will enhance it.  However, reducing it too much
will result in a more pronounced anharmonic region and so the solitons will be prone to acceleration and dissipation will occur in a shorter timescale.

This behavior should also be compared to that of a moving soliton in the left well, when the part of the condensate trapped in the right well does not contain a soliton.
Unlike the case of a dark soliton in a `closed' harmonic trap geometry, the presence of a region for the emitted sound to escape leads to a slow decay; the envelope of such anti-damped oscillations are
shown by the blue lines in Fig. 2 (bottom). The comparison to this scenario 
 demonstrates clearly that the presence of the soliton in
the right well acts so as to stabilize the left soliton, enabling it to survive/oscillate for much longer times.

\section{Experimental considerations: Double/Multiple Wells}

We seek to identify a directly observable effect that could be measured in experiments.
Despite the appealing feature of the `identity-reversal' of the two solitons in the
idealized cut-off harmonic trap discussed previously, such an effect is unlikely to be directly observed
in an experiment; this is due to the sensitivity of solitons to the locally strongly anharmonic region near the
barrier separating the two regions in a realistic experimental double-well geometry.
As an alternative, and to make direct links with experiments,
 we therefore consider the simpler scenario of looking at changes in the motion
of a soliton in one well, due to the presence, or absence, of a soliton in the neighbouring well.

Thus, we now discuss experimental geometries in which solitons have been successfully generated, namely Gaussian traps and optical lattice geometries.
For simplicity we restrict this study to identical solitons.


\subsection{Gaussian induced, double-well potential}

\begin{figure}[b]
\centering{
\includegraphics[scale=0.32]{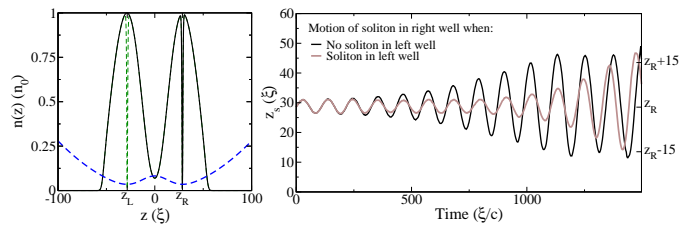}
}
\caption{(Color Online)  Left:  Density profile (black and grey dashed) and trapping potential (lower dashed blue) of an atomic condensate in a double-well potential, with either a
single $0.1c$ soliton in the right well ($z_R=28.5\xi$, black lines), or also a second identical soliton located in the symmetric position of the left well ($z_L=28.5\xi$, green dashed).
Right:  Position of the soliton in the right well when it is the only soliton present in the trap (black) and when there is another
soliton in the left well (grey). 
(Parameters: $V_0/\mu =0.97$, $d=14.6 \xi$.)}
\label{gauss}
\end{figure}

We first consider a double well geometry of the form $V(z) = m \omega_z^2 z^2/2 + V_0e^{{-z^2}/{2d^2}}$;
such a geometry can be formed by adding a repulsive Gaussian dimple to a harmonic trap, e.g. by the addition of a blue
detuned laser~\cite{bluelaser}, with  $V_0$ and $d$ characterizing the height and width of the
Gaussian.  As before, two solitons are placed at the peak densities of the two wells spanned by the condensate, with their respective profiles shown in Fig.~\ref{gauss} (Left).

The wells have been chosen such that
the effective harmonic frequency of each well
is of the order of the longitudinal frequency $\omega_z$ used earlier;
however, {\it within each well}, the pronounced anharmonicity which
arises from the lack of symmetry
leads to the decay of the soliton. Hereafter we scale distances to
the healing length $\xi$ ($\xi \approx l_z/4.9$ for the previous parameters).

Given the rapid decay of a soliton in such an asymmetric trap, we instead investigate
%
%
the motion of a soliton in the right well depending on whether there is a soliton in the well to the left or not.  
We find that the addition of a
soliton in the left well stabilizes the other soliton to a large extent, as shown in  Fig.~\ref{gauss} (Right); as a result, instead of quickly decaying, it continues oscillating 
with roughly the same amplitude (i.e. same energy/depth)
for a timescale of the order of $800 \xi/c$, 
before eventually decaying. We also note a slight
phase difference between the two cases.

\subsection{Optical double-well potential}

To increase the soliton lifetimes we next probe a regime in which each well of the trap is itself symmetric, but also not harmonic.
Intersecting laser beams at an angle can be used to generate an optical lattice
or an array of traps of variable depth and periodicity.
This configuration is modelled by $V(z) = V_0 \mbox{cos}^2 \left({2 \pi z}/{d} \right)$, with the lattice parameters $V_0$ and $d$ characterizing the lattice's depth and spacing respectively.  Again,
solitons of equal speed are initially placed at the peak condensate densities.

\begin{figure}[t]
\centering{
\includegraphics[scale=0.3]{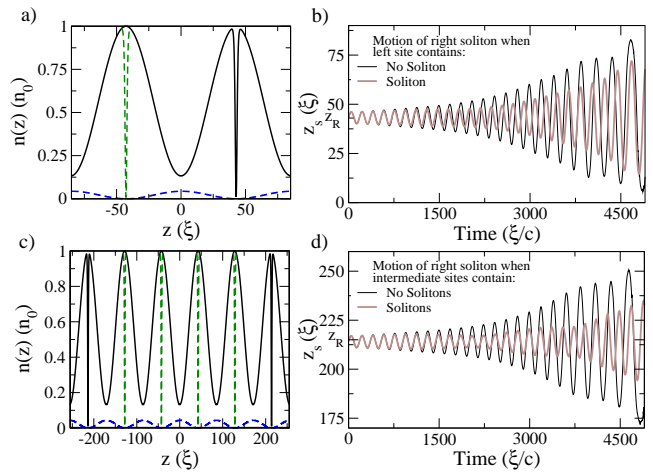}
}
\caption{(Color Online) Left: Density profiles for
a condensate spanning (a) two, or (c) six sites 
of an array of optical traps.
In both cases, 
 the right site contains a
$v_s=0.1c$ soliton (black). In addition,
in (a) the left site may
or may not contain an identical soliton (green dashed),
whereas in (d) the leftmost site also contains an identical soliton,
while the condensate in the intermediate sites may or may not contain solitons.
 Right:  Temporal evolution of the soliton in the right well 
showing clearly that the presence of a soliton in neighbouring wells has a stabilizing effect:
(b) Comparison of case when there is no soliton in the
left well (black) to when this site contains a soliton (grey).
(d) Comparison of the dynamics of the soliton in the rightmost trap,
when there are no intermediate
solitons (black) and when the intermediate sites contain identical solitons (grey). 
(Parameters: $V_0/\mu = 0.87$ and $d=170.9\xi$).  }
\label{lattice}
\end{figure}

As before, we initially consider a periodic array of traps of relatively large periodicity \cite{birkl}, in which the condensate however only spans two wells, as shown in Fig.~\ref{lattice} (top).
We investigate the evolution of an initially propagating soliton ($v=0.1c$) in one well, when the condensate in the other well does not contain a soliton (black line);
this is contrasted to the case when there is an identical soliton in the other well (grey line).
It is apparent that the
presence of the other soliton leads to a slower decay.  This indicates that the soliton in the right well absorbs the sound created by the motion of the
soliton through the inhomogeneous background of the left well.  The more symmetric nature of the trap results in the soliton experiencing a slower dissipation than in the Gaussian, double-well
potential of Fig.~\ref{gauss}.

\subsection{Array of optical traps}

Finally, we wish to 
explicitly demonstrate the role of
dark solitons as both {\it absorbers} and {\it emitters} of 
sound~\cite{Leadbeater}, by investigating
the effect of adding further solitons to a condensate 
spanning a number of lattice sites greater than two.

%

We consider a condensate spanning 6 lattice sites, and place two identical slowly-moving solitons in the wells at each end of the condensate, as shown in Fig. 4 (bottom). The soliton in the right well
is found to decay (black lines in Fig. 4(d)) on a timescale of $\sim 25$ oscillations, due to the partial escape of the emitted sound to the other wells, while the sound from the soliton in the
leftmost well arrives too late to stabilize it. This should be contrasted to the case when each intermediate site contains solitons of the same depth. In that case, the solitons in the neighbouring
wells clearly have the net effect of partially stabilizing the end soliton against decay over a prolonged period, as evident by the grey lines in Fig. 4(d).

Faster solitons were found to decay on a more rapid timescale, as the time required for the oscillating soliton to probe the anharmonic region of the trap decreases rapidly with increasing soliton speed. However, this effect should be observable in carefully controlled experiments, and it does not require identical soliton speeds.

\section{Conclusions}

In conclusion, we have 
demonstrated that two or more solitons propagating in an inhomogeneous medium can interact and exchange energy without ever coming into direct contact,
or colliding with each other; the physical mechanism of this exchange is based on the mutual exchange of sound energy over distances much larger than the soliton size. 
One could also interpret this process as follows:
one soliton perturbs the background condensate, setting up a dipole mode in it; this dipole mode, perturbs the condensate region where a second soliton is confined, thereby forcing the other (propagating) soliton to see a temporally-varying background condensate; this leads to a change in the interaction of the second soliton with the background, which can periodically lead to absorption of energy from the background, rather than emission.
In an idealized geometry, this interaction leads to a complete energy transfer which results in periodic identity reversal of the two solitons.

Under current experimental conditions this identity reversal may not be observable, as more oscillations are required than have been observed up until now~\cite{Becker,Weller,Stellmer}, and one would additionally need to significantly minimize the region where the potential becomes anharmonic.  
However, experiments should be able to detect that the presence of a second soliton in a condensate confined in an appropriate double well geometry has a stabilizing effect.  
The experiment we have in mind requires an experimental set up which can reproduce the same initial conditions to good accuracy, so that differences in individual soliton trajectories in one well after a number of 10-15 oscillations arising from the presence of another soliton located in a spatially separated well can be observed (e.g. after expansion imaging).
Generalizing this to an array of traps (which could be thought of as a soliton gas \cite{El})
the addition of more solitons was found to lead to a slower dissipation of the outer solitons.
Extensions of this work could include the interplay between soliton-sound interaction and Josephson
effects \cite{andy_martin}, a deeper investigation of the role of the intermediate states in
a multi-site configuration in terms of resonant tunnelling \cite{mark_fromhold,Schlagheck},
and the role of quantum and thermal fluctuations.

We would like to thank K. Bongs, S.L. Cornish, D.M. Gangardt and N.G. Parker for discussions.  We acknowledge funding from the UK EPSRC.


\begin{thebibliography}{99}
\bibitem{solitons} R.K. Dodd, J.C. Eilbeck, J.D. Gibbon, and H.C. Morris. \emph{Solitons and Nonlinear Wave Equations} (London: Academic, 1982).
\bibitem{kivshar_review} Y. S. Kivshar and B. Luther-Davies, Phys. Rep.
{\bf{298}}, 81 (1998).
\bibitem{frantzeskakis_book}\emph{Emergent Nonlinear Phenomena in Bose-Einstein Condensates: Theory and Experiment,} edited by P. G. Kevrekidis, D. J. Frantzeskakis, and R. Carretero-Gonz{\'{a}}lez, Springer Series on Atomic, Optical, and Plasma Physics Vol. 45 (Springer, Berlin, 2008).
\bibitem{burger} S. Burger, K. Bongs, S. Dettmer, W. Ertmer, K. Sengstock, A. Sanpera, G.V. Shlyapnikov, and M. Lewenstein, Phys. Rev. Lett. {\bf{83}}, 5198 (1999).
\bibitem{phillips}J. Denschlag, J.E. Simsarian, D.L. Feder, C.W. Clark, L.A. Collins, J. Cubizolles, L. Deng, E.W. Hagley, K. Helmerson, W.P. Reinhardt, S.L. Rolston, B.I. Schneider, and W.D. Phillips,
Science {\bf{287}}, 97 (2000).
\bibitem{Becker}
C. Becker, S. Stellmer, P. Soltan-Panahi, S. D\"{o}rscher, M. Baumert, E.M. Richter, J. Kronj\"{a}ger, K. Bongs, and K. Sengstock, Nat. Phys. {\bf{4}}, 496 (2008).
\bibitem{Weller}
A. Weller, J.P. Ronzheimer, C. Gross, J. Esteve, M.K. Oberthaler, D.J. Frantzeskakis, G. Theocharis, and P.G. Kevrekidis, Phys. Rev. Lett. {\bf{101}}, 130401 (2008).
\bibitem{Stellmer} S. Stellmer, C. Becker, P. Soltan-Panahi, E.M. Richter, S. D\"{o}rscher, M. Baumert, J. Kronj\"{a}ger, K. Bongs, and K. Sengstock,
Phys. Rev. Lett.  {\bf{101}}, 120406 (2008).
\bibitem{huletsols} D. Dries, S.E. Pollack, J.M. Hitchcock, and R.G. Hulet, Phys. Rev. A {\bf{82}} 033603 (2010).  
\bibitem{Tik} V. Tikhonenko, J. Christou, B. Luther-Davies, and Y.S. Kivshar, 
Opt. Lett. {\bf{21}}, 1129, (1996).
\bibitem{dutton} Z. Dutton, M. Budde, C. Slowe, and L.V. Hau, Science {\bf{293}}, 663 (2001).
\bibitem{engelsdyna} P. Engels and C. Atherton, Phys. Rev. Lett, {\bf{99}},
160405 (2007).
\bibitem{ketterlemerging} G.B. Jo, J.H. Choi, C.A. Christensen, T.A. Pasquini, Y.R. Lee, W. Ketterle, and D.E. Pritchard, Phys. Rev. Lett. {\bf{99}}, 240406 (2007).
\bibitem{KZ}
B. Damski and W. H. Zurek, Phys. Rev. Lett. {\bf 104}, 160404 (2010).
%
\bibitem{busch_anglin} 
Th. Busch and J.R. Anglin, Phys. Rev. Lett. {\bf{84}}, 2298 (2000).
\bibitem{makarov} G. Huang, M. G. Velarde, and V. Makarov, Phys. Rev. A {\bf{64}}, 013617 (2001).
\bibitem{frantzeskakis2002} D.J. Frantzeskakis, G. Theocharis, F.K. Diakonos, P. Schmelcher, and Y. S. Kivshar, Phys. Rev. A {\bf{66}}, 053608 (2002).
\bibitem{Leadbeater}
N.G. Parker, N.P. Proukakis, M. Leadbeater and C.S. Adams,
 Phys. Rev. Lett. {\bf{90}}, 220401 (2003).
\bibitem{pelinovsky} D.E. Pelinovsky, D.J. Frantzeskakis, and P.G. Kevrekidis, Phys. Rev. E. {\bf{72}}, 016615 (2005).
\bibitem{Parker2010}
N.G. Parker, N.P. Proukakis and C.S. Adams,
Phys. Rev. A {\bf{81}}, 033606 (2010).
\bibitem{muryshev}
A.E. Muryshev, H.B. van Linden van den Heuvell, and G.V. Shlyapnikov,
Phys. Rev. A {\bf{60}}, R2665 (1999).
\bibitem{fedichev} P.O. Fedichev, A.E. Muryshev, and G.V. Shlyapnikov, Phys Rev. A {\bf{60}}, 3220 (1999).
\bibitem{Carr}
L.D. Carr, M.A. Leung, and W.P. Reinhardt, J. Phys. B {\bf{33}}, 3983 (2000).
\bibitem{Feder}
D.L. Feder, M.S. Pindzola, L.A. Collins, B.I. Schneider, and C.W. Clark, Phys. Rev. A {\bf{62}}, 053606 (2000).
\bibitem{jbrand}
J. Brand and W.P. Reinhardt, Phys. Rev. A {\bf{65}}, 043612 (2002).
\bibitem{mury} 
A.E. Muryshev, G.V. Shlyapnikov, W. Ertmer, K. Sengstock, and M. Lewenstein, 
Phys. Rev. Lett. {\bf{89}}, 110401 (2002).
\bibitem{sacha}
J. Dziarmaga, Z. P. Karkuszewski, K. Sacha, J. Phys. B {\bf 36}, 1217 (2003).
%
\bibitem{proukakis_analogies} 
N.P. Proukakis, N.G. Parker, D.J. Frantzeskakis and  C.S. Adams, 
J. Opt. B: Quantum Semiclass. {\bf{6}}, S380 (2004). 
\bibitem{schmelcher} G. Theocharis, P. Schmelcher, M.K. Oberthaler, P.G. Kevrekidis, and D.J. Frantzeskakis, Phys. Rev A {\bf{72}}, 023609 (2005).

\bibitem{jackson}
  B. Jackson, N.P. Proukakis, and C.F. Barenghi, Phys. Rev. A {\bf{75}}, 051601(R) (2007).
  B. Jackson, C.F. Barenghi, and N.P. Proukakis, J. Low Temp. Phys. {\bf{148}}, 387 (2007).
\bibitem{cockburn}
  S.P. Cockburn, H.E. Nistazakis, T.P. Horikis, P.G. Kevrekidis, N.P. Proukakis, and D.J. Frantzeskakis, 
Phys. Rev. Lett. {\bf{104}}, 174101 (2010).
\bibitem{ruostekoski} 
  A. D. Martin and J. Ruostekoski, Phys. Rev. Lett. {\bf{104}}, 194102 (2010).
\bibitem{quantumdecay} D.M. Gangardt and A. Kamenev, Phys. Rev. Lett. {\bf{104}}, 190402 (2010).
\bibitem{foursa} D. Foursa and P. Emplit, Phys. Rev. Lett. {\bf{77}}, 4011 (1996).
\bibitem{frantzeskakis_condmat} G. Theocharis, A. Weller, J. P. Ronzheimer, C. Gross, M. K. Oberthaler, P. G. Kevrekidis, D. J. Frantzeskakis 
Phys. Rev. A {\bf{81}}, 063604 (2010).
\bibitem{segev}
C. Rotschild, B. Alfassi, O. Cohen, and M. Segev, 
Nat. Phys. {\bf{2}}, 769 (2006).
\bibitem{parkerjphys}
N.G. Parker, N.P. Proukakis, M. Leadbeater and C.S. Adams, 
J. Phys. B {\bf{36}}, 2891 (2003).
\bibitem{parametricdriving} N.P. Proukakis, N.G. Parker, C.F. Barenghi and C.S. Adams, 
Phys. Rev. Lett. {\bf{93}}, 130408 (2004).
\bibitem{bluelaser} M.R. Andrews, C.G. Townsend, H.J. Miesner, D.S. Durfee, D.M. Kurn, and W. Ketterle, 
Science {\bf{275}}, 637 (1997).
\bibitem{birkl} G. Birkl, F.B.J. Buchkremer, R. Dumke, and W. Ertmer, Optics Communications. {\bf{191}}, 67 (2001).
\bibitem{El}
G.A. El and A.M. Kamchatnov, Phys. Rev. Lett. {\bf{95}} 204101 (2005).
\bibitem{andy_martin} A.M. Martin, Private Communication.
\bibitem{mark_fromhold} M. Fromhold, Private Communication.
\bibitem{Schlagheck}  C. Eltschka and P. Schlagheck, Phys. Rev. Lett.  {\bf{94}}  014101 (2005).


\end{thebibliography}
\end{document}